\pdfoutput=1
\documentclass[a4paper,11pt]{article}
\usepackage{jcappub} 
\usepackage{lineno}

	\newcommand{\bq}{\begin{equation}}
	\newcommand{\eq}{\end{equation}}
	\newcommand{\bqn}{\begin{eqnarray}}
	\newcommand{\eqn}{\end{eqnarray}}
	\newcommand{\nb}{\nonumber}

\arxivnumber{1234.56789} 
\title{Enhanced Blandford Znajek Jet in Loop Quantum Black Hole}

\author[a]{Hong-Xuan Jiang}
\author[a]{Indu K. Dihingia}
\affiliation[a]{Tsung-Dao Lee Institute, Shanghai Jiao Tong University, Shengrong Road 520, Shanghai, 201210, China}
\author[b,c,d]{Cheng Liu}
\affiliation[b]{School of Physics and Astronomy, Shanghai Jiao Tong University, 800 Dongchuan Road, Shanghai, 200240, China}
\affiliation[c]{Shanghai Frontiers Science Center for Gravitational Wave Detection, 800 Dongchuan Road, Shanghai 200240, China}
\affiliation[d]{United Center for Gravitational Wave Physics (UCGWP),  Zhejiang University of Technology, Hangzhou, 310023, China}
\author[a,b]{Yosuke Mizuno}
\author[d,e]{Tao Zhu}
\affiliation[e]{Institute for Theoretical Physics \& Cosmology, Zhejiang University of Technology, Hangzhou, 310023, China}

\emailAdd{hongxuan\_jiang@sjtu.edu.cn, ikd4638@gmail.com, liuc09@sjtu.edu.cn, mizuno@sjtu.edu.cn, zhut05@zjut.edu.cn}

\abstract{The Blandford-Znajek (BZ) process powers energetic jets by extracting the rotating energy of a Kerr black hole. It is important to understand this process in non-Kerr black hole spacetimes. In this study, we conduct two-dimensional and three-dimensional two-temperature General Relativistic Magnetohydrodynamic (GRMHD) simulations of magnetized accretion flows onto a rotating Loop-Quantum black hole (LQBH). Our investigation focuses on the accretion flow structure and jet launching dynamics from our simulations. 
We observe that the loop quantum effects increase the black hole angular frequency for spinning black holes.
This phenomenon intensifies the frame-dragging effect, leading to an amplification of the toroidal magnetic field within the funnel region and enhancement of the launching jet power. 
It is possible to fit the jet power following a similar fitting formula of the black hole angular frequency as seen in the Kerr black hole. 
Based on the General Relativistic Radiation Transfer (GRRT) calculation, we find that the jet image from LQBH has a wider opening angle and an extended structure than the Kerr BH.}

\begin{document}
\maketitle
\flushbottom

\section{Introduction} \label{sec:intro}
Tests on the theory of general relativity have been implemented since it was discovered in 1905. It is still the most accepted theory of gravity. Recently, the horizon-scale observation of supermassive black holes (SMBH) M\,87$^*$ and Sgr\,A$^*$ by Event Horizon Telescope Collaboration (EHTC) unveiled the physics of the black holes (BH) and the accretion flow surrounding them with the highest angular resolution that we can achieve on Earth \citep{EventHorizonTelescope:2019dse,EHT2022a}. Although we still cannot distinguish Kerr BH and BH solutions from other theories of gravity \citep[e.g.,][]{EHT2022f,Afrin2023,LQG_BH,2023arXiv231204288J, 2020PhRvL.125n1104P, Ozel2022,Younsi2023,2023CQGra..40p5007V}, an upper limit for the ``charge" of the BH can be measured \citep{Kocherlakota2021,EHT2022f}. Within the acceptable parameter space for the SMBHs, there are still many unique features from alternative theories of gravity. One of them is the structure of the jets and formation mechanisms in non-GR frameworks.

Relativistic jets have been discovered in astrophysical systems, from stellar mass BHs (Gamma-ray Burst, X-ray binary, etc.) to SMBHs (active galactic nuclei (AGNs)) \citep[e.g.,][]{1992ApJ...397L...5M, 1999ApJ...519L..17S, 2006csxs.book..381F,2019ARA&A..57..467B}. Particularly,  due to the proximity of the radio galaxy M87, meticulous observations of its jet have revealed edge-brightening and precession phenomena \citep{2018A&A...616A.188K, Lu2023, Cui2023}. These observations provide important information about the jet launching mechanism and frame-dragging of the SMBH. Blandford-Znajek (BZ) process is believed to be the possible mechanism for jet launching \citep{Blandford1977, Yang:2024kpz}. In this mechanism, energy is extracted from the spin of the black hole to power the relativistic jets. 
A pertinent inquiry would be whether the BZ process remains functional within the framework of alternative theories of gravity. There are several analytical investigations for the BZ process in the alternative theory of gravity \cite[e.g.,][]{Pei2016,Banerjee2021,2023arXiv230706878C}.

GRMHD simulations of stringy black holes have been implemented in several works \citep{Mizuno2018, Roeder2023, 2023arXiv231020040C}. In \cite{2023arXiv231020040C}, they demonstrated that a stronger jet is generated via the BZ process in stringy rotating black holes. 
Using the parameterized Johannsen-Psaltis (JP) metric \cite{Johannsen:2013szh}, energy extraction from the spacetime with deviation from GR has investigated through the BZ process in \cite{Chatterjee:2023rcz}.
This indicates that the BZ process still works as the dominant energy extraction mechanism for creating jets in non-Kerr spacetimes.
Although the BZ process is the most promising mechanism to extract the energy for the black hole, there are several other possible mechanisms such as superradiance \citep{Jha:2022tdl, Khodadi:2022dyi, Brito:2015oca} and magnetic reconnection \citep{Khodadi:2023juk,Comisso:2020ykg} via the Penrose-like process \cite{Penrose:1971uk}.
When the oscillation frequency $\omega_R$ of perturbations to a Kerr black hole is less than the product of the perturbation azimuthal number $m$ and the event horizon angular velocity $\Omega_+$, i.e., $\omega_R<m\Omega_+$, these perturbations are amplified by the rotation of the black hole to the extent that the energy radiated to infinity can exceed that of the initial perturbation \citep{1971JETPL..14..180Z, 1972JETP...35.1085Z, Starobinsky:1973aij}. It is a so-called superradiance process. 
Similarly, magnetic reconnection is a quite common phenomenon in magnetized accretion flow which occurs the condition with anti-parallel magnetic fields \cite{Nathanail2020, Dihingia:2021ncv, Ripperda:2021zpn, Jiang2023, Jiang2024, Selvi:2024lsh}. 
In a magnetically arrested disk (MAD), magnetic reconnection happens in the equatorial plane. It forms a current sheet that connects to the magnetosphere of the BH, in which plasmoids are generated through tearing instability \cite{Ripperda:2021zpn, Jiang2023}. In the process, plasmoids that contain negative energy fall into the horizon while others obtain energy from the rotation of the BH. In some extreme conditions, energy extraction through reconnection is even higher than the BZ process \cite{Khodadi:2023juk}.

Loop quantum gravity (LQG) is another important alternative theory of gravity that adeptly addresses classical singularities, including those associated with the Big Bang and black holes \citep{LQG_BH, Liu:2020ola,2023arXiv231204288J}. 
The recently proposed self-dual spacetime is a regular static spacetime metric obtained by the mini-superspace approach based on the polymerization procedure in the LQG \citep{LQG_BH}, which has no curvature singularity and whose regularity is determined by two parameters of the LQG: the minimal area and the Barbero-Immirzi parameter. The corresponding rotating LQBH was obtained in Ref. \citep{Liu:2020ola} using the modified Newman-Janis procedure, which brought the study of the LQBH to a wide range of astrophysical tests. Till date, the black holes have been subjected to several intensive observational tests, such as quasi-periodic oscillations in X-ray high precision timing of black hole binaries \citep{Liu:2023vfh}, the highest resolution images of the M87* central black hole \citep{Liu:2020ola, 2023arXiv231204288J} and the central black hole Sgr A* of the Milky Way \citep{2023arXiv231204288J}, and decades of monitoring of the orbit of the S2 star near the Sgr A* \citep{Yan:2022fkr}, see e.g., \citep{Zhu:2020tcf, Liu:2021djf, Daghigh:2020fmw, Bouhmadi-Lopez:2020oia, Fu:2021fxn, Brahma:2020eos} and references therein.

In our previous work by \cite{2023arXiv231204288J}, we implemented a large parameter survey to study the possible range of parameters under the current limit of EHT observations. In general, the Event Horizon Telescope (EHT) constraints favor a relatively high BH spin ($a \gtrsim 0.5$), and the LQG parameter $P$ is constrained to be $P\lesssim0.15$ for Sgr\,A$^*$ and M\,87$^*$. 
In this work, the emission from a geometrically thick accretion flow \cite{Chen:2021lvo} was taken into account.
Another constraint of the polymeric function $P$ from direct comparison with theoretical calculation and the observed shadow radius is done by \cite{2023CQGra..40p5007V}. It puts an upper limit of $P\lesssim0.05$.
Note that the physical properties of the accretion flow are still not fully understood. These comparisons with EHT observation may be influenced by the emission from the gaseous accretion disk \cite{Gralla:2020pra} and it is important to understand it. This work is the first step in this direction. Since the astrophysical jet is one aspect people have been observing for many decades in different classes of astrophysical black holes (from stellar mass to supermassive \cite{Sari:1999mr, Blandford:2018iot}). 
Therefore, in this work, we study the BZ process using the same LQBH metric as \cite{2023arXiv231204288J} within the possible parameter space. This study aims to explore the repercussions of the LQG effect on both the accretion process and the magnetic field surrounding black holes. Additionally, using the GRRT calculation, we further study the image of jets in LQBH.

This paper is organized as follows: In section 2, we introduce our numerical method and setup, We present our result in section 3, and finally, in section 4 we provide our conclusions.

\section{Numerical Method}

In this work, we perform a series of two-temperature GRMHD simulations of magnetized tori in a loop quantum black hole spacetime (see Appendix~\ref{sec:LQBH} for detail) using the $\tt BHAC$\footnote{\url{https://bhac.science/}} code \citep{Porth2017, Olivares2019}. $\tt BHAC$ code solves the ideal GRMHD equations in geometric units ($G_{\rm N}M = c = 1$ and $1/\sqrt{4\pi}$ is included in the magnetic field\footnote{Our LQBH metric corrects the gravitational constant by a factor related to the polymeric parameter $P$. To distinguish LQG and Newtonian gravitational constants, we label them by $G_{\rm LQG}$ and $G_{\rm N}$. (See Appendix~\ref{sec:LQBH} for more detail.)}), they are given by:
\begin{equation}
	\nabla_{\mu}\left(\rho u^{\mu}\right)=0,\,\,\,
	\nabla_{\mu}T^{\mu\nu}=0,\,\,\,
	\mathrm{and}\,\,\,
	\nabla_{\mu}^*F^{\mu\nu}=0,
\end{equation} 
where $\rho$ is the rest-mass density, $u^{\rm \mu}$ is the fluid four-velocity, $T^{\rm \mu\nu}$ is the energy momentum tensor, and $^*F^{\rm \mu\nu}$ is the dual of the Faraday tensor. For the details of the numerical methods, readers may refer to \citet{Porth2017, Olivares2019}. 

The simulations are initialized with a hydrodynamic equilibrium torus given by \cite{Font2002}. Following \cite{Mizuno2018}, all the simulations we run have the same $r_{\rm{in}} = 10.3 \,r_{\rm g}$, where $r_{\rm g}\equiv G_{\rm N}M/c^2$ is the gravitational radius of the black hole. By adjusting the specific angular momentum, we let the total mass of the tori with different black hole parameters be the same for consistency. Thus, we make a fair comparison of plasma dynamics around a black hole with different dimensionless spin parameters $(a)$ and polymeric function parameter $(P)$. The simulation domain has a radius of $r=2\,500\,r_{\rm g}$. The spatial resolution of the 2D simulations is $1\,024\times 512$ in Kerr-Schild-like coordinates (see Appendix~\ref{sec:LQBH}). The 3D simulations are implemented with Static Mesh Refinement (SMR) with an effective resolution of $384\times192\times192$. An ideal-gas equation of state with a constant adiabatic index of $\Tilde{\Gamma} = 4/3$ is adopted throughout the simulations. 

A pure poloidal magnetic field is initially provided in the torus to initiate the magneto-rotational instability in all cases. It is done by setting up a non-zero toroidal component for the initial vector potential, which is given as follows:
\begin{equation}
    A_{\rm \phi} \propto \rho/\rho_{\rm max}-0.25.
    \label{Eq: SANE}
\end{equation}
All other components of the vector potential are set to zero. The strength of the magnetic field is initially fixed by setting the minimum of plasma $\beta$ ($\beta_{\rm min}=100$). This ensures a relatively weak magnetic field, forming accretion flows in a Standard and Normal Evolution (SANE) regime.

Based on the flow properties from 3D GRMHD simulations, General Relativistic Radiation Transfer (GRRT) calculation is performed with the {\tt RAPTOR}\footnote{\url{https://github.com/jordydavelaar/raptor}} code. We modified the spacetime of the code following our previous semi-analytical GRRT work \citep{2023arXiv231204288J}. Non-thermal electron distribution function (eDF) is included with the kappa eDF (details see Appendix~\ref{sec:nonthermal}). The grid resolution of the GRRT calculations is $1\,000\times1\,000$ pixels, which covers a field of view (FoV) of roughly $1.5\times 1.5\,\rm mas^2$ ($400\times400\,\rm r_{\rm g}^2$). We use M\,87$^*$ as the target, whose mass and distance are $6.5\times10^9\,\rm M_{\odot}$ and $16.8\,\rm Mpc$ \citep{EventHorizonTelescope:2019dse}. The mass unit is determined by fitting the total flux at $230\,\rm GHz$ to $1\,\rm Jy$.

\begin{figure*}
    \centering    
    \includegraphics[height=0.41\linewidth]{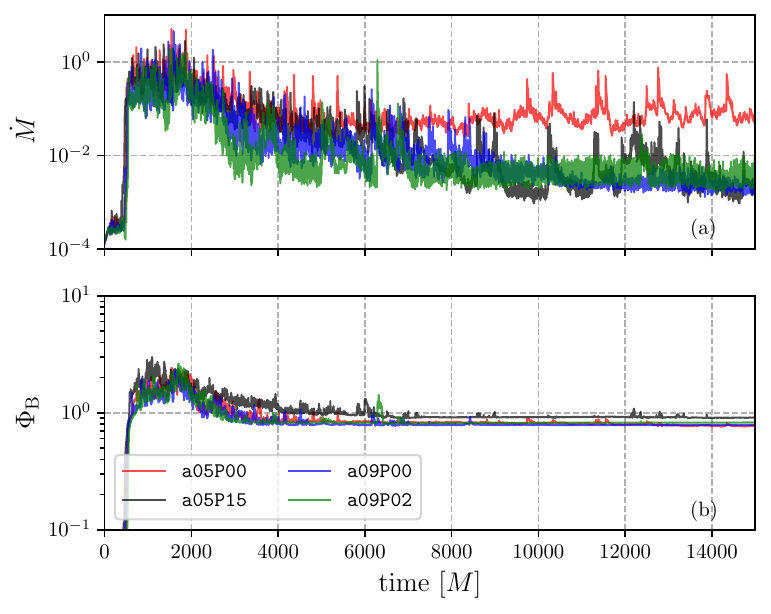}
    \includegraphics[height=0.41\linewidth]{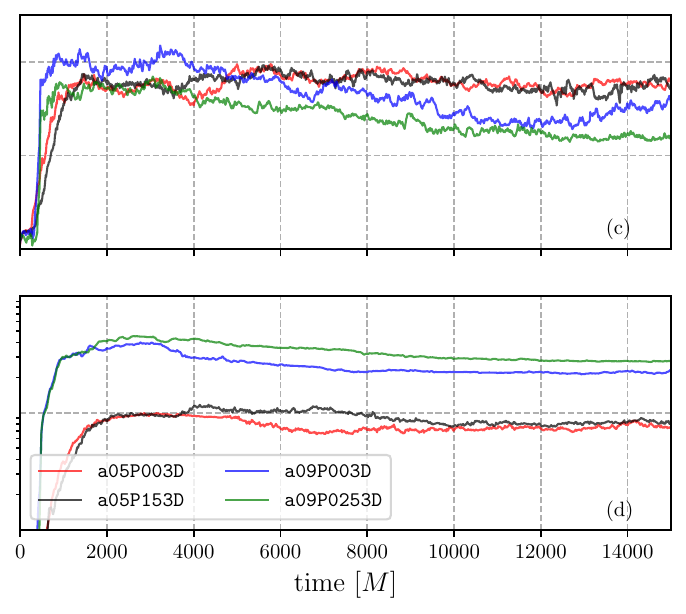}
    \caption{The left two panels (a) and (b) are the mass accretion rate and magnetic flux evolutions from 2D GRMHD simulations {\tt a05P00}, {\tt a05P15}, {\tt a09P00}, and {\tt a09P02}. The right two panels (c) and (d) are from the 3D cases {\tt a05P003D}, {\tt a05P153D}, {\tt a09P003D}, and {\tt a09P0253D}.}     \label{fig: Mdot}
\end{figure*}

\section{Result}
\subsection{Accretion flows surrounding an LQBH}

The LQBH parameters, BH spin $a$ and LQG polymeric function parameter $P$, of the simulations are selected considering a black hole solution of the LQBH metric (see table~\ref{table: LQBH}). The upper limits of these parameters are set to avoid naked singularity \citep{Liu:2020ola}. We label the GRMHD simulations by their $a$ and $P$ values. The 3D cases are indicated with an extra $\tt 3D$. Here, we focus our discussions primarily on the cases with $a=0.5$ and $a=0.9$. We present the accretion rate $\dot{M}$ as well as the magnetic flux $\Phi_{\rm B}$ of these cases in the upper and lower panels of Fig.~\ref{fig: Mdot}, respectively. We compute the mass accretion rate and dimensionless magnetic flux rate measured at the event horizon \citep{Porth2019}, which are given by:
\begin{equation}
\begin{aligned}
    &\dot M = \int_0^{2\pi}\int_0^{\pi} \rho u^r \sqrt{-g}d\theta d\phi, \\
    &\Phi_{\rm B} = \frac{1}{2}\int_0^{2\pi}\int_{0}^{\pi}\left|B^r\right|\sqrt{-g}d\theta d\phi. \label{Eq: B-flux}
\end{aligned}
\end{equation}

In Fig.~\ref{fig: Mdot}(a) and (b), we see that after simulation time $t>6,000\,\rm M$, the accretion rate and magnetic flux for all the cases reach a quasi-steady state. The quasi-steady value of the accretion rate decreases with the increase in the values of $a$ and $P$. For instance, the time-averaged accretion rate for the 2D runs with $a=0.5$, $P=0$, and $0.15$, is $\langle \dot{M} \rangle =0.062,$ and $0.005$, respectively. A similar reduced trend in accretion rate also happens in 3D cases. For runs with $a=0.9$, a lower accretion rate is seen in the LQBH than in the Kerr BH. Unlike the 2D results, the magnetic flux in the 3D simulations gets higher when $a$ and $P$ increase, which can be attributed to the more complex dynamo process in the 3D simulations \citep[e.g.,][]{2023arXiv231100034J}.

\begin{figure*}[t]
    \centering
    \includegraphics[height=0.41\linewidth]{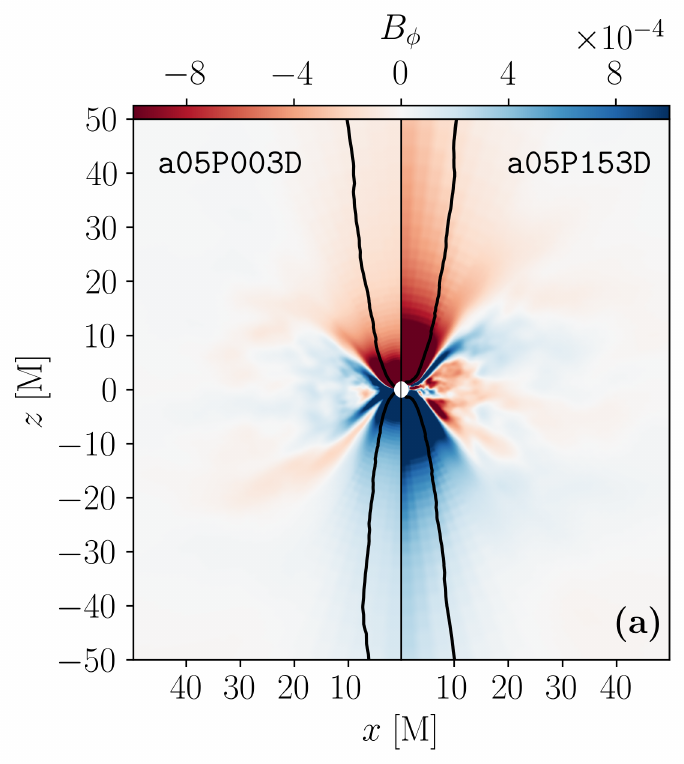}
    \includegraphics[height=0.41\linewidth]{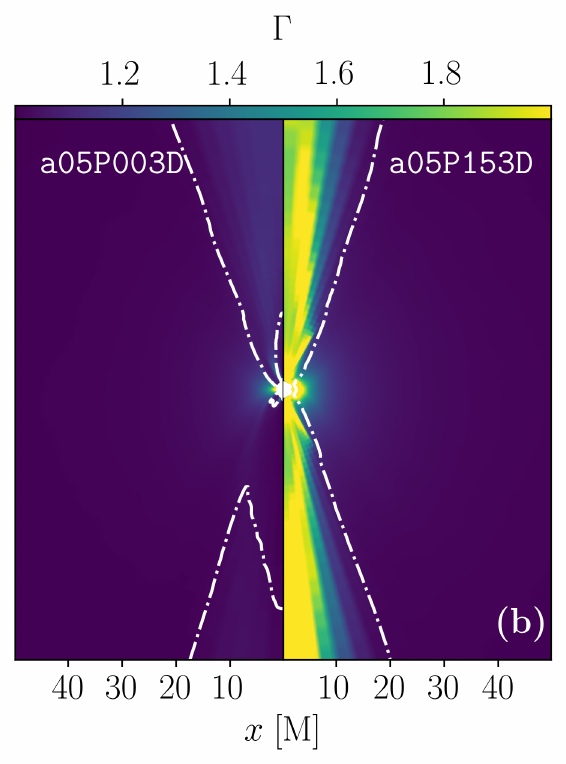}
    \includegraphics[height=0.41\linewidth]{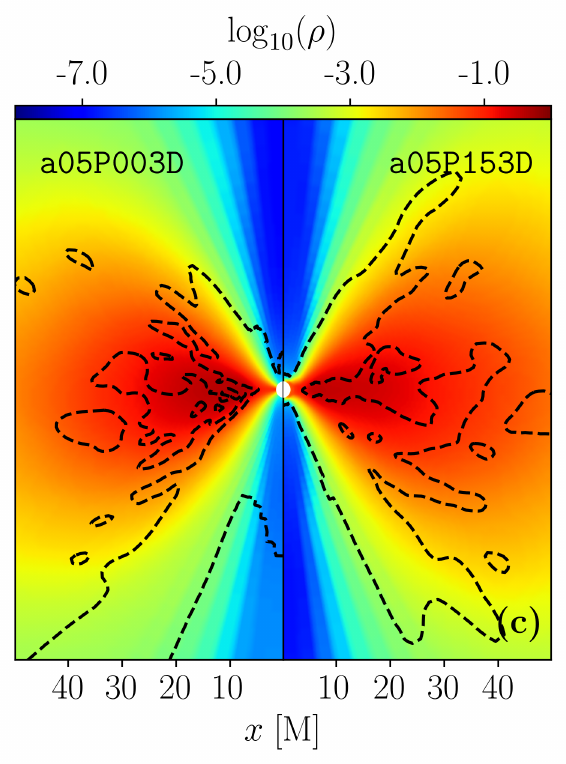}
    \caption{Comparisons between the 3D GRMHD results of LQBH and Kerr BH. In each panel, the left half is Kerr BH with $a=0.5$, while the right half is from LQBH with $a=0.5$ and $P=0.15$. The panels (a), (b), and (c) show the toroidal magnetic field $(B_{\rm \phi})$, Lorentz factor $(\Gamma)$, and mass density $(\rho)$ distribution, respectively. The contour in panels (a), (b), and (c) correspond to the boundaries of $\sigma=1$ (black), $-hu_t=1$ (white dash-dotted), and $u^r=0$ (black dashed), respectively.}
    \label{fig: LQBH_compare}
\end{figure*}

We use the 3D simulation results to compare the flow structure around Kerr BH and LQBH, showing the distribution of time averaged (a) toroidal magnetic field, (b) Lorentz factor, and (c) density between LQBH and Kerr BH in Fig.~\ref{fig: LQBH_compare}. 
The left half of each panel represents the case {\tt a5P003D} while the right half is {\tt a5P153D}. Case {\tt a5P153D} has a $P$ value of $0.15$, which is a relatively large value, close to the critical value ($P_{\rm max}=0.1957$). The white contour in panels (a), (b), and (c) corresponds to the boundaries of magnetization $\sigma=b^2/\rho=1$, $-hu_t=1$, and $u^r=0$, respectively. The toroidal magnetic field is amplified via the frame-dragging effect from rotating BHs \citep[e.g.,][]{Jiang2023}. Stronger toroidal magnetic field and slightly larger jet region ($\sigma>1$) are observed from case {\tt a05P153D} than {\tt a05P003D} (see Fig.~\ref{fig: LQBH_compare}(a)). A stronger magnetic field near the horizon leads to a decreased accretion rate with increasing $P$ (see Fig.~\ref{fig: Mdot}(a)).
A higher Lorentz factor in the jet is seen in case {\tt a05P153D} than {\tt a05P003D}. Meanwhile, a higher Lorentz factor also appears in the region that is close to the horizon ($r<10\,\rm r_{\rm g}$) near the equatorial plane. The faster-rotating plasma close to the LQBH is primarily responsible for it. From the $-hu_t=1$ contour, we observe a smaller area for gravitationally bound regions in the LQBH than that of the Kerr black hole. Lastly, the density (shown in panel Fig.~\ref{fig: LQBH_compare}(c)) is a little higher in the funnel region of Kerr BH, which means it has less magnetization. A unique feature of the accretion flow surrounding LQBH is that the inflow tunnel is thicker than Kerr BHs. This suggests that the geodesic motions in Kerr and LQBH are not the same. The $u^r=0$ contour reveals that the outflow surface for case {\tt a05P15} is much closer to the horizon than case {\tt a05P15}.

In conclusion, the LQ effect strengthens the magnetic field in the funnel region, leading to a lower accretion rate and a stronger jet. In the following sections, we point out that this effect is a result of the higher angular frequency near the horizon of LQBH.

\subsection{Jet launching from LQBH}\label{sec:enhanced_spin}

The BZ process, which extracts power from a rotating black hole, is thought to produce powerful relativistic jets \citep{Blandford1977}. To investigate the BZ process in LQBH, it is necessary to compute the angular frequency at the event horizon, i.e.,
\begin{equation}
    \Omega_{\rm H} = \left(-\frac{g_{t\phi}^{\rm BL}}{g_{\phi\phi}^{\rm BL}}\right)_{r_{\rm H}},
\end{equation}
where $r_{\rm H}$ is the radius of the outer horizon.
For the case of LQBH, it is reduced to
\begin{equation}
    \Omega_{\rm H}^{\rm LQBH} = \\
    \Omega^{\text{Kerr}}_{\text{H}}\Bigg(1+\frac{2(a^2+a^4+2\sqrt{1-a^2}-2)}{a^2 (1-a^2)}P + O(P^2)\Bigg). \label{Eq:Omega_H} 
\end{equation}
For Kerr BH, the angular frequency is given by $\Omega_{\rm H}^{\rm Kerr} = a/2r_{\rm H}^{\rm Kerr}$ \citep{Tchekhovskoy2010}. Eq.~(\ref{Eq:Omega_H}) reveals a positive correlation of angular frequency at the horizon for LQG. Therefore, even if BH spin is the same, with positive $P$, LQBH has a higher angular frequency.
 
The relation between the angular frequency of a BH and jet power from GRMHD simulations is discussed in \cite{Tchekhovskoy2010}. They provided a relation between the jet power of Kerr BH to the magnetic flux and the angular frequency at the horizon as:
\begin{equation}
    P^{\rm BZ}\approx k \Phi_{\rm tot}^2\Omega_{\rm H}^2,\label{Eq:BZpower}
\end{equation}
where $\Phi_{\rm tot}$ gives the total magnetic flux in the jet. 
\begin{figure*}
    \centering
    \includegraphics[height=0.33\linewidth]{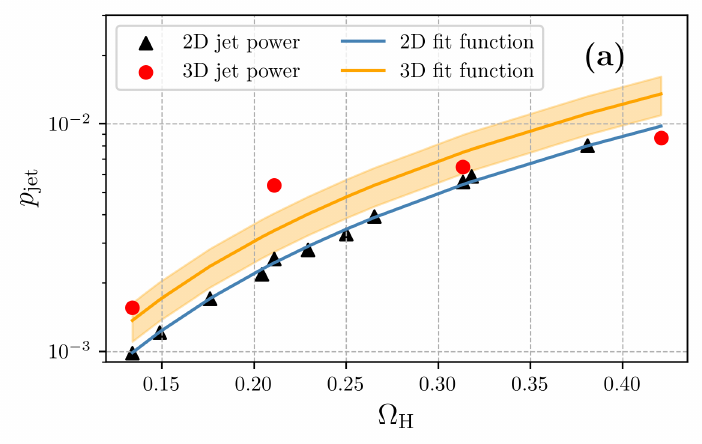}
    \includegraphics[height=0.33\linewidth]{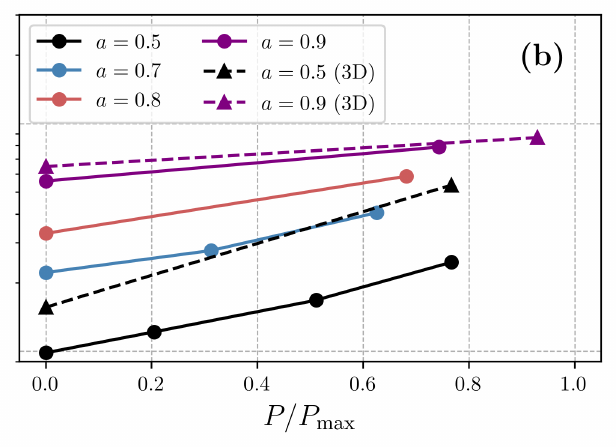}
    \caption{Panel (a): the distribution of normalized time-averaged ($t=13,000-15,000\,$M) BZ jet powers ($p_{\rm jet}$) for different simulation models. The fitting function of the jet power is shown by the solid lines (Eq.~\ref{Eq:BZpower}). The shaded region represents the standard deviation (STD) of the fitting parameter $k$. Panel (b): jet power evolution with fixed BH spin $a$ and increasing $P$. Different colors of the lines represent LQBHs with different spins. The solid lines represent the 2D simulation result, and the dashed lines are from 3D simulation.} \label{fig:BZ_fit}
\end{figure*}

To study the jet power quantitatively, we first define the boundary of the jets by the contour that magnetization $\sigma=1$. Jet power can be measured by calculating the energy flux that passes through the spherical surfaces within the funnel region ($\sigma\geq1$) at $50\,\rm r_g$ \citep{Nathanail2020,EHT2022e}. Accordingly, the jet power is expressed as:
\begin{equation}
    P_{\rm jet}=\int_0^{2\pi}\int_{\theta_{\rm jet}}(-T_t^r-\rho u^r)\sqrt{-g}d\theta d\phi,
\end{equation}
where the range of $\theta_{\rm jet}$ covers the funnel region ($\sigma\geq1$). 

To understand how the jet power changes with BH angular frequency, we plot the normalized BZ jet power ($p_{\rm jet}=P_{\rm jet}/\Phi^2_{\rm jet}$) obtained from our simulations in LQBHs with spinning BH as black triangles (2D) and red (3D) dots (see in Table~\ref{table: LQBH}) in Fig.~\ref{fig:BZ_fit}(a). The jet powers are calculated from the time-averaged value of $p_{\rm jet}$ from $t=13,000$ to $15,000\,\rm M$. We fit Eq.~\ref{Eq:BZpower} to the data for 2D and 3D simulations from them individually, which are overlayed with the blue (2D) and orange (3D) curves in the figure. Since the BZ mechanism dominates the power of the jet ($P_{\rm jet}/P_{\rm em}\sim 1$, where $P_{\rm em}$ is Poynting flux dominated jet power), we consider $P_{\rm jet}$ to be approximately $P_{\rm BZ}$. The best-fit parameter from the 2D simulations is obtained as $k=0.055\pm 0.0006$. However, from the 3D result, we find $k=0.076\pm 0.0147$ with larger uncertainty (see Fig.~\ref{fig:BZ_fit}(a)). The shaded region around the solid lines shows the uncertainty in the values of the fitting parameters. For 2D cases, the uncertainty is much lower than in 3D cases; therefore, it is not visible in the figure. The higher value of uncertainty in 3D cases is due to the limited number of simulation models. There are differences between the $k$ values we got from our 2D and 3D GRMHD simulations and those obtained from the force-free axisymmetric simulations in \cite{Tchekhovskoy2010} and the original analytical work by \cite{Blandford1977}. 
The cause of this difference can be attributed to the influence of high temperatures in the jet within our GRMHD simulations and the inertial effects of plasma.

To better understand the jet power enhancement by including of LQG effect, we plot the jet power as a function of $P$ with fixed BH spins in Fig.~\ref{fig:BZ_fit}(b). Different colors represent different values of BH spin $a$. As seen in Eq.~\ref{Eq:Omega_H}, the angular frequency at the event horizon $\Omega_H$ becomes faster in increase of $P$. As predicted from Eq.~\ref{Eq:BZpower}, a positive correlation is seen between polymeric function $P$ and jet power across different black hole spins. For example, case {\tt a05P153D} ($P/P_{\rm max}\sim 0.78$) has roughly four times higher jet power than that of {\tt a05P003D}.

\begin{figure}
    \centering
    \includegraphics[width=.6\linewidth]{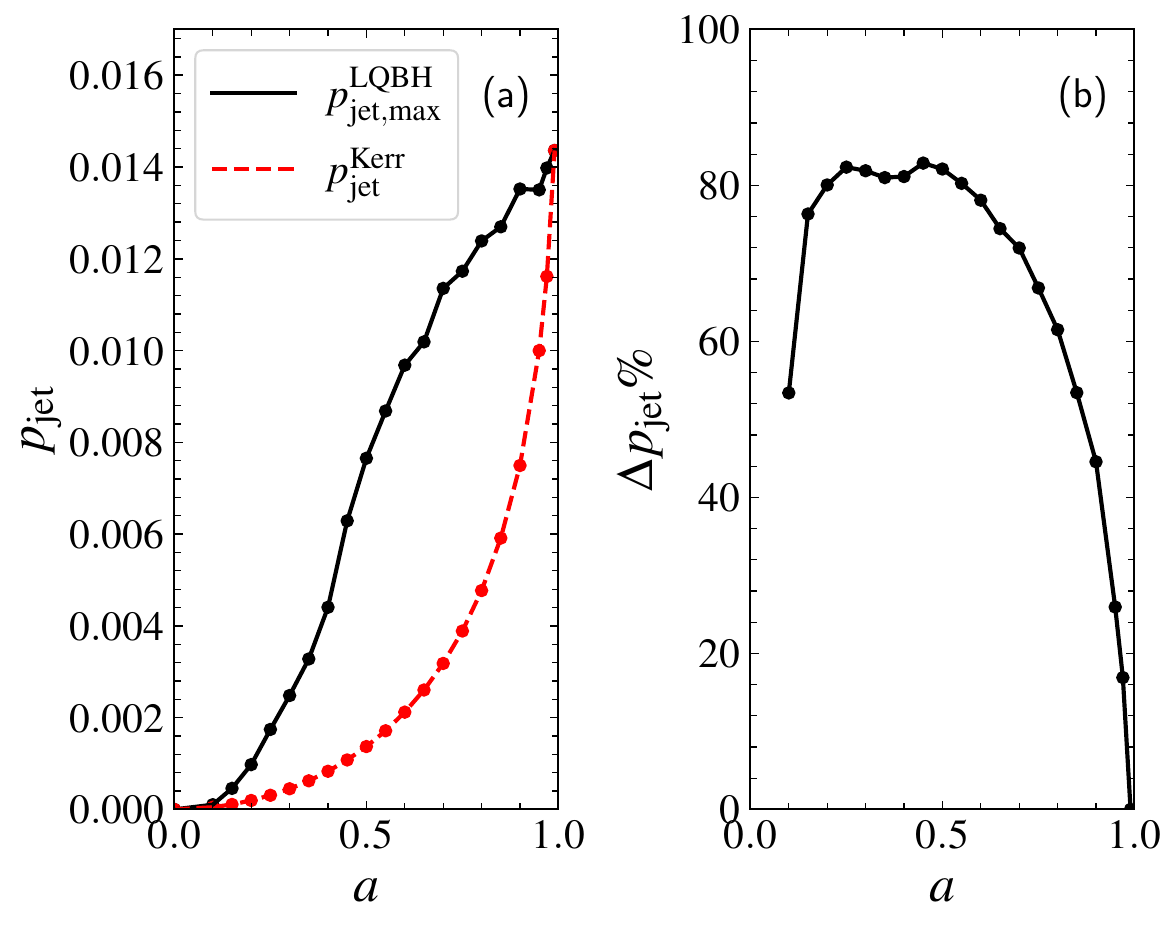}
    \caption{Panel (a): plot of maximum jet power from LQBH ($p^{\rm LQBH}_{\rm jet,max}$, black solid line) and for Kerr BH ($p^{\rm Kerr}_{\rm jet}$, red dashed line) for different Kerr parameters. Panel (b): plot of the relative increase in jet power for LQBH as compared Kerr BH ($\Delta p_{\rm jet}$) }     \label{fig:Pjetminmax}
\end{figure}

We should mention that all of the GRMHD simulations we perform are within the SANE limit, which contains a relatively weaker and less ordered magnetic field.
However, \cite{Akiyama2019,EHT2022e} show the accretion flows of Sgr\,A$^*$ and M\,87$^*$ are unlikely to be SANE regimes. The MAD model is more favored by both sources. A significant contrast between MAD and SANE accretion flows lies in the strength of the magnetic field. In MAD scenarios, the magnetic field near the horizon is strong enough against the gravitational pull from the BH at certain disk locations. 
This phenomenon halts the accretion process temporarily. A strong and ordered magnetic field is observed in Sgr~A$^*$ in the recent polarimetric observation results \cite{Akiyama_2024a,Akiyama_2024b}. 
MAD is currently the most favored one. However, even MAD can not pass all the observational constraints (especially the variability constraint) made by EHT \cite{Akiyama_2024a, Akiyama_2024b}.
Thus, the accretion flow models of Sgr~A$^*$ are still under debate \cite{Jiang2023, Jiang2024, Ressler:2020voz, Liska:2017alm}. New accretion flow models are required to address these issues, e.g., \cite{S:2023gta, Jiang2024, Dihingia:2024dsk}. For the case of M\,87$^*$, more observational evidence pointed out it is likely to be MAD \cite{Yuan:2022mkw}. 
Recently quasi-periodic jet precession of M\,87 is observed \cite{Cui2023}. However, current simulations of tilted MAD flow indicated that the persistent jet precession cannot generate \cite{Chatterjee:2023ber}.

In MAD flow, the dynamics of the accretion flow are strongly influenced by the magnetic field \cite{Ripperda:2021zpn, Zhang:2023hgh, Begelman:2021ufo}. Flux eruptions drive powerful wind with power up to $10\%$ of the accretion power \cite{Chatterjee:2022mxg}, which is positively related to BH spin \cite{Yang:2021ndp}. The relatively high magnetization in the wind also makes it hard to distinguish from the jet. How much energy from the BZ process is injected into the wind is still not clear. Flux eruptions also impact the stability of the jet, leading to fluctuations in jet power \cite{Tchekhovskoy:2011zx}. Another point is that the jet from MAD flow contains stronger magnetic pressure than the gas pressure \cite{Jiang2023}, which should contribute to the jet power as well. These effects add more uncertainty in distinguishing the contribution purely from the magnetic field or the electromagnetic BZ process. On the contrary, SANE models do not show such eruption events, and we observe a steady jet throughout. Accordingly, the safest way to study the influences of black hole parameters on BZ jets is to consider a weakly magnetized SANE model. In this work, we aim to study the observable effect of LQG, leaving the more complicated plasma and magnetic field effects in MAD flow for our future work.

\subsection{Jet power variations with LQBH parameters}

\begin{figure*}
    \centering    
    \includegraphics[height=0.45\linewidth]{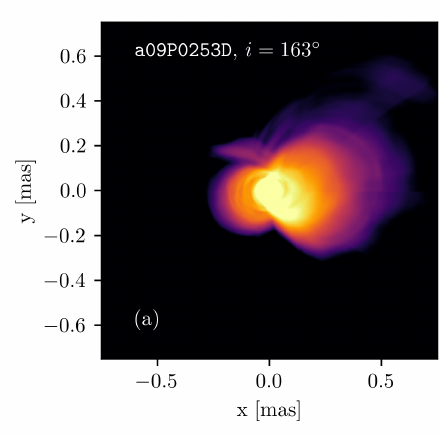}
    \includegraphics[height=0.45\linewidth]{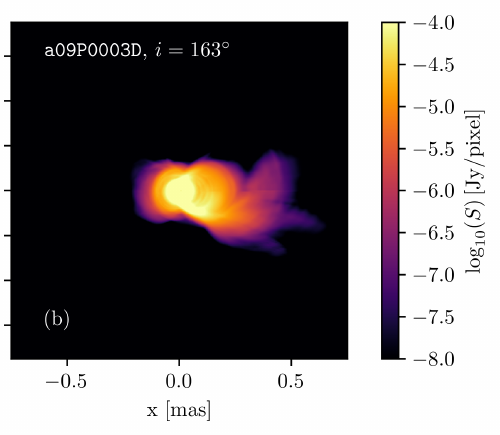}
    \caption{Panel (a) and (b) present the averaged GRRT images at $86\,\rm GHz$ of the jets from case {\tt a09P0253D} and {\tt a09P003D}. Average ranges from $14\,000$ to $\,\rm 15\,000\,\rm M$. }     \label{fig:Jet}
\end{figure*}

In the previous sections, we have shown the enhanced BZ jet from a LQBH as compared to a Kerr BH. Maximum $P$ value depends on the BH spin in LQBH.
In Fig.~\ref{fig:Pjetminmax}(a), we present the maximum of normalized BZ power for LQBH $p^{\rm LQBH}_{\rm jet, max}=P_{\rm jet}|_{\rm max}/\Phi_{\rm jet}^2$ as a function of the BH spin (the solid curve). We also plot the normalized jet power for a Kerr BH ($P=0$) case in the same figure (a red dashed line). We calculate the value of $P_{\rm jet}$ following Eq.~\ref{Eq:BZpower}, where we use our best-fit parameter from 3D simulations. In Fig.~\ref{fig:Pjetminmax}(b), we show the relative increase in jet power for LQBH as compared with Kerr BH for a given BH spin, i.e., $\Delta p_{\rm jet}=(p^{\rm LQBH}_{\rm jet, max}-p^{\rm Kerr}_{\rm jet})/p^{\rm LQBH}_{\rm jet, max}\times100\%$. 
From the figure, in both scenarios, the power extracted via the BZ mechanism consistently increases with the spin of the BH. However, the increase in LQBH is notably faster compared to Kerr BH. The jet power tends to converge at both extremes of spin ($a\rightarrow0$ and $a\rightarrow 1$). It implies that the maximum power attainable through the BZ process from a LQBH is constrained by the maximum extractable power from a Kerr BH without surpassing it. However, with spin parameter $a\sim0.5$, the BZ jet from LQBH can be more than $80\%$ stronger than the jet from Kerr BH. Moreover, it reveals that the same jet power could be the result of combinations of different values of $a$ and $P$.

Considering the correlation between jet power and BH spin, particularly in the context of M\,87$^*$, where a BH spin value around $a \approx 0.9$ is favored \cite{Akiyama2019, Tamburini:2019vrf}, the potential amplification of jet power from LQG may be restricted. Similarly, this limitation applies to Sgr~A$^*$ if the BH spin is relatively low ($a \lesssim 0.1$), as suggested by \cite{Fragione:2020khu}. However, subsequent observations by the EHT suggest a relatively high spin for Sgr~A$^*$ ($a \gtrsim 0.5$) \cite{EHT2022e, Akiyama_2024a, Akiyama_2024b}. The conclusion of high BH spin for M~87$^*$ and Sgr~A$^*$ is made under the assumption of Kerr spacetime. In the framework of LQBH, energy extraction can reach a similar magnitude with Kerr, but at a lower BH spin value ($a\lesssim0.7$) \cite{2023arXiv231204288J}.

\subsection{Extended jet image for LQBH}
In the previous sections, we found that the LQG effect enhances the BZ process. In the same spin parameter, we observed higher magnetization and Lorentz in the jet region in LQBH as compared to Kerr BH (see Fig.~\ref{fig: LQBH_compare}). The magnetic energy is injected into the kinetic energy in different ways, e.g., through turbulence \citep{Meringolo2023} and reconnection \citep{Ball2018}. Both of them generate a power law distribution of the electrons, in which the slope of the distribution function depends on the microphysics of the plasma. Combining the Maxwell-J\"{u}ttner and power law distribution (see Appendix~\ref{sec:nonthermal} for detail), we adopt the kappa eDF for the nonthermal emission in our GRRT calculation, which forms bright jet emission \citep{2018A&A...612A..34D}. To show the impact of LQG on the images of the jet, we present the milliarcsecond GRRT images at $86\,\rm GHz$ in Fig.~\ref{fig:Jet}. Panels (a) and (b) represent images from {\tt a09P003D} and {\tt a09P0253D}.
The BH spins are identical for both cases in this figure, but case {\tt a09P0253D} has an extreme LQG effect that produces a more powerful jet. In GRRT images, this case also has a stronger and more extended emission structure, resulting in a bright edge-brightening jet with a wider jet opening angle.

\section{Summary and discussion}

In this study, using GRMHD and GRRT simulations, we explored the effect of loop quantum gravity on the accretion process around LQBHs. We conducted a range of 2D and 3D GRMHD simulations using the LQBH metric with different values of spin parameters $(a)$ and polymeric functions $(P)$. The summaries of our investigation are presented below.

\begin{enumerate}
    \item In GRMHD simulations, our study highlights the significant impact of the polymeric function $(P)$ in the realm of LQBH, particularly near the event horizon. Increasing $P$ leads to a noteworthy decrease in accretion rate,
    attributed to heightened magnetic pressure from a stronger dynamo process fueled by the increased angular frequency of LQBH. Additionally, a positive correlation is observed between $P$ and both the Lorentz factor and the power of the BZ jet.
    
    \item We fit angular frequency at the horizon to study BZ jet power across different LQBH parameters based on our GRMHD simulations.
    Our GRMHD simulations show a small difference in parameter $k$ compared to the value in \cite{Tchekhovskoy2010}. This is because our simulations have more complicated physics than the axisymmetric force-free simulations in \cite{Tchekhovskoy2010}. Notably, the fitting parameter from 3D simulations is slightly higher than the 2D one, suggesting deviation from the axisymmetric assumption.
    
    \item By comparing the BZ power extracted from a Kerr BH with the maximum power extracted for LQBH, we found that for the intermediate spin parameter range for LQBH, the BZ jet power could be almost double that of the BZ power of Kerr BH. However, for the extreme limits ($a\rightarrow0~{\rm and}~a\rightarrow1$), the power extracted for LQBH and Kerr BH are the almost same.
    
    \item The stronger frame-dragging effect in LQBH amplifies the magnetic field in 
    the funnel region, which generates more non-thermal emissions than the Kerr BH. In GRRT images at 86\,GHz, the jet in the LQBH creates a more extended structure than that of the Kerr BH.
\end{enumerate}

Finally, we discuss some of the caveats of our study. Our study considers only SANE accretion models. However, the accretion flow surrounding Sgr\,A$^*$ and M\,87 is known to be closer to MAD \citep[e.g.,][]{Akiyama2019,EHT2022e,Yuan2022,2022NatAs...6..103C}, which we will conduct in our future work. Our current simulation focuses on the simplest single-loop SANE simulation to understand the impact of LQG on accretion flow and jet. The higher-order effects, e.g., hot spots, light curve variation, etc., are all ignored. On top of this, studying the spectral (e.g., image, spectra) and timing properties (e.g., variability) of astrophysical sources demands a two-temperature radiative cooling accretion flow model and different magnetic field configurations \citep[e.g.,][]{Mizuno2021, Dihingia2023, Jiang2023, Jiang2024}. 
We will carry forward such studies in the future.

\appendix

\section{LQBH spacetime}\label{sec:LQBH}

The self-dual spacetime arises from the symmetry-reduced model of LQG, which corresponds to homogeneous spacetimes. Such spacetime is geodesically complete and free from any spacetime curvature singularity at the center of the BH. The metric in the Boyer-Lindquist coordinates is given by \cite{LQG_BH,Liu:2020ola,2023arXiv231204288J},
 \begin{equation}
     d s^2 = \frac{\Delta}{\Sigma}(dt-a\sin^2\theta d\phi)^2-\frac{\Sigma}{\Delta}dr^2-\Sigma d\theta^2 -\frac{\sin^2\theta}{\Sigma}(a dt-(k^2+a^2)d\phi)^2,  
      \label{metric}
 \end{equation}
 with
\begin{eqnarray}
\Delta= \frac{(r-r_+)(r-r_-)r^2}{(r+r_*)^2}+a^2,
\Sigma=k^2(r)+a^2\cos^2\theta,\
k^2= \frac{r^4+a^2_0}{(r+r_*)^2},
\end{eqnarray}
where $a$ is the specific angular momentum (spin parameter) of the black hole. $r_+=2 G_\text{LQG} M/(1+P)^2$ and $r_{-} = 2G_\text{LQG} M P^2/(1+P)^2$ are the two horizons for the corresponding spherical LQBH, and $r_{*}= \sqrt{r_+ r_-} = 2G_\text{LQG} MP/(1+P)^2$ with $G_\text{LQG}$ representing the gravitational constant in the LQBH, $M$ denoting the ADM ({Arnowitt-Deser-Misner}) mass of the solution, and $P$ being the polymeric function, which is given by
\bqn
P \equiv \frac{\sqrt{1+\epsilon^2}-1}{\sqrt{1+\epsilon^2}+1},   \label{P_epsilon}
\eqn 
where $\epsilon$ denotes a product of the Immirzi parameter $\gamma$ and the polymeric parameter $\delta$, i.e., $\epsilon=\gamma \delta \ll 1$. The parameter $a_{0}$ is defined as
\bqn
a_0 = \frac{A_{\rm min}}{8\pi},
\eqn
where $A_{\rm min}$ represents the minimum area gap of LQG. It is interesting to mention that $A_{\rm min}$ is related to the Planck length $l_{\rm Pl}$ through $A_{\min} \simeq 4 \pi \gamma \sqrt{3} l_{\rm Pl}^2$ \citep{Sahu:2015dea}. Thus, $a_0$ is proportional to $l_{\rm Pl}$ and is expected to be negligible. Hence, phenomenologically, the effects of $a_0$ on spacetime are expected to be very small, and in this paper, we will only focus on the effects on the scale of a few to thousands of Schwarzschild radii of LQG and set $a_0=0$. The gravitational constant in the rotating LQBH ($G_\text{LQG}$) can be related to the Newtonian gravitational constant ($G_{\rm N}$) by 
\bqn
G_\text{LQG}=G_{\rm N}\frac{(1+P)^2}{(1-P)^2}.
\eqn  
Finally, the event horizon for rotating LQBH can be obtained by solving $\Delta(r_{h\pm})=0$. The explicit expression of $r_{h\pm}$ can be found in \cite{Liu:2020ola}. Note that, for $P=0$, $r_{h\pm}$ transitions to the widely recognized expression for the Kerr black hole, i.e., $r_{h\pm}=M \pm \sqrt{M^2-a^2}$.

In our numerical simulations, we substitute the Newtonian gravitational constant into our metric (Eq.~(\ref{metric})) to make a fair comparison between LQBH and Kerr BH {(i.e., $G=G_{\rm LQG}$)}. Considering that dimensionless spin $a$ is also proportional to the gravitational constant, which also needs to be scaled by a factor of ${(1+P)^2}/{(1-P)^2}$. We should note that \cite{Afrin2023} did not scale the gravitational constant to the Newtonian one, which makes the comparison between LQBHs with different polymeric functions inconsistent. Therefore, the result of our work is expected to be different from \cite{Afrin2023}.

Another parameter, the Immirzi parameter $\gamma$, has a lot of choices from different considerations \citep[see][]{BenAchour:2014qca, Frodden:2012dq, Carlip:2014bfa, Taveras:2008yf}. It has been shown that its value can even be a complex number \citep{Frodden:2012dq, BenAchour:2014qca, Carlip:2014bfa}, or considered as a scalar field in which the value would be fixed by the dynamics \citep{Taveras:2008yf}. We adopt the commonly used value $\gamma = 0.2375$ from the black hole entropy calculation \citep{Meissner:2004ju}. 

We use Kerr-Schild coordinates to avoid a singularity at the event horizon. Accordingly, the coordinate transformations from Boyer-Lindquist-like coordinate to Kerr-Schild-like coordinate are given by,
\bqn
dr_\text{KS}&=&dr, \\
d\theta_\text{KS}&=&d\theta, \\
dt_\text{KS}&=&dt - \frac{2M\mathbf{A} r}{\Delta}dr_\text{KS}, \\
d\phi_\text{KS} &=& d\phi -\frac{a}{\Delta}dr_{KS}.
\eqn
With this transformation, the metric in the new coordinate is obtained as follows \citep{Kocherlakota2023},
\bqn
ds^2 = g^{\rm KS}_{tt}dt^2 + 2g^{\rm KS}_{tr}drdt + 2g^{\rm KS}_{t\phi}dtd\phi \nb
+ g^{\rm KS}_{rr}dr^2 + g^{\rm KS}_{r\phi}drd\phi + g^{\rm KS}_{\theta\theta}d\theta^2 +g^{\rm KS}_{\phi\phi}d\phi^2,
\eqn
with
\bqn
g^{\rm KS}_{tt}&=&-\frac{\Delta-a^2 \sin^2\theta}{\Sigma},  \\
g^{\rm KS}_{tr}&=&\frac{1}{\Sigma}[\mathbf{A} (2Mr+a^2 \sin^2\theta)-a^2\sin^2\theta], \\
g^{\rm KS}_{t\phi}&=&-\frac{a \sin^2\theta (k^2+a^2-\Delta)}{\Sigma},\\
g^{\rm KS}_{rr}&=&\frac{\mathbf{A}}{\Sigma}[\mathbf{A}(\Delta+4Mr+a^2\sin^2\theta)-2a^2\sin^2\theta], \\
g^{\rm KS}_{r\phi}&=&-\frac{a\sin^2\theta}{\Sigma}[(r^2+a^2)\mathbf{A}^2+2Mr\mathbf{A}-a^2 \sin^2\theta],\\
g^{\rm KS}_{\theta\theta}&=&\Sigma, \\
g^{\rm KS}_{\phi\phi}&=&-\sin^2\theta\Bigg[\frac{a^2 \sin^2\theta\Delta-(k^2+a^2)^2}{\Sigma}\Bigg],
\eqn
where
\bqn
\mathbf{A}=\frac{k^2+a^2}{r^2+a^2}.
\eqn
{\bf Note that ${\bf A}\rightarrow 1$, as $r \rightarrow \infty$.} 
\section{LQBH parameters}\label{sec:LQBH_parameter}

The parameters and labels of the GRMHD simulations are listed in Table~\ref{table: LQBH}. The first row is the 2D cases. The cases in the second row are labeled with an extra {\tt 3D}, marking they are 3D simulations.

\begin{table*}[]
\caption{List of the BH spins ($a$) and the value of the Polymeric function ($P$) for GRMHD models.
}
\centering
\begin{tabular*}{1.0\textwidth}{@{\extracolsep{\fill}}lllllllllll}
\hline
\multicolumn{1}{l}{case} & ${\tt a5P00}$ & ${\tt a5P10}$ & ${\tt a5P15}$ & ${\tt a7P00}$ & ${\tt a7P03}$ & ${\tt a7P06}$ & ${\tt a8P00}$ & ${\tt a8P04}$ & ${\tt a9P00}$ & ${\tt a9P02}$ \\ 
\hline
$a$         & 0.5             & 0.5           & 0.5           & 0.7           & 0.7           & 0.7           & 0.8           & 0.8           & 0.9           & 0.9           \\
$P$         & 0.0             & 0.10          & 0.15          & 0.0          & 0.03          & 0.06          & 0.0          & 0.04          & 0.0           & 0.02         \\ 
\hline
\end{tabular*}
\begin{tabular*}{0.6\textwidth}{@{\extracolsep{\fill}}lllll}
\hline
\multicolumn{1}{l}{case} & ${\tt a5P003D}$ & ${\tt a5P153D}$ & ${\tt a9P003D}$ & ${\tt a9P0253D}$ \\ \hline
$a$        & 0.5           & 0.5           & 0.9           & 0.9 \\
$P$        & 0.0           & 0.15          & 0.0          & 0.025        \\ 
\hline
\end{tabular*}
\label{table: LQBH}
\end{table*}

\section{Non-thermal electron distribution function}
\label{sec:nonthermal}
The GRRT calculation in this work contains two components, thermal and nonthermal synchrotron emission. The thermal synchrotron emission is calculated with the electron temperature obtained from our two-temperature GRMHD simulations \cite{Mizuno2021, Jiang2023}. The non-thermal contribution is calculated using the kappa eDF:
\begin{equation}
    \frac{dn_{\rm e}}{d\gamma_{\rm e}}=\frac{N}{4\pi}\gamma_{\rm e}\sqrt{\gamma_{\rm e}^2-1}\left(1+\frac{\gamma_{\rm e}-1}{\kappa w}\right)^{-(\kappa+1)},
\end{equation}
where $n_{\rm e}$ is the electron number density, $\gamma_{\rm e}$ is the electron Lorentz factor, $\kappa$ is the slope of the distribution function of the non-thermal particles and $N$ is a normalization factor (see \cite{Pandya2016} for detail). The microphysics of particle acceleration is included with a sub-grid method, which adopts the fitting formula from Particle-In-Cell (PIC) simulations of special-relativistic turbulence \citep{Meringolo2023}:
\begin{equation}
        \kappa(\beta,\sigma)=2.8+\frac{0.2}{\sqrt{\sigma}} + 1.6\sigma^{-6/10}\tanh{\left(2.25\beta\sigma^{1/3}\right)}. \label{Eq: kappa_tur}
\end{equation}

\acknowledgments
Y.M. is supported by the Shanghai Municipality orientation program of Basic Research for International Scientists (Grant No.\,22JC1410600), the National Natural Science Foundation of China (Grant No.\,12273022), and the National Key R\&D Program of China (No.\,2023YFE0101200). 
T.Z. is supported in part by the National Key Research and Development Program of China under Grant No.\,2020YFC2201503, the National Natural Science Foundation of China under Grant No.\,12275238 and No.\,11675143, the Zhejiang Provincial Natural Science Foundation of China under Grant No.\,LR21A050001 and LY20A050002,  and the Fundamental Research Funds for the Provincial Universities of Zhejiang in China under Grant No.\,RF-A2019015.
The simulations were performed on TDLI-Astro, Pi2.0, and Siyuan Mark-I at Shanghai Jiao Tong University.




\end{document}